\documentstyle[preprint,aps]{revtex}
\begin{document}
\title{Incomplete melting of the Si(100) surface from 
  molecular-dynamics simulations using the Effective-Medium
  Tight-Binding model. }

\author{K. Stokbro,$^{1,2,3}$  K. W. Jacobsen,$^3$ J. K. N\o rskov,$^3$
D. M. Deaven,$^4$ C. Z. Wang,$^4$ and K. M. Ho$^4$}
\address{
$^{1}$Mikroelektronik Centret, Danmarks Tekniske Universitet, 
Bygning 345\o , DK-2800 Lyngby, Denmark. \\
$^2$ Scuola Internazionale Superiore di Studi Avanzati, via Beirut 4,
I-34014 Trieste, Italy. \\
$^3$Center for Atomic-scale Materials Physics and Physics Department, \\
Danmarks Tekniske Universitet, DK 2800 Lyngby, Denmark \\
$^4$ Ames Laboratory, Ames, Iowa 50011}

\maketitle

\begin{abstract}
  We present molecular-dynamics simulations of the Si(100) surface in
  the temperature range 1100-1750K. To describe the total energy and
  forces we use the  Effective-Medium Tight-Binding
  model. The defect-free surface is found to melt at the bulk melting
  point, which we determine to be 1650 K, but for a surface with dimer
  vacancies we find a  pre-melting of the first two
  layers 100 K below the melting point.  We show that these findings
  can rationalize recent experimental studies of the high temperature
  Si(100) surface.
\end{abstract}

\section{Introduction}
Most work on the Si(100) surface has focussed on the behaviour at low
and intermediate temperatures, while to our knowledge there have to
date only been two experimental studies \cite{MeWo93,FrFeCo94a} of the
atomic structure at temperatures near the bulk melting point. Both of
these studies indicate that the surface undergoes a phase transition
at a temperature below the bulk melting point, however, there is no
consistent picture of the nature of the transition, the exact
transition temperature, and the atomic structure beyond the phase
transition.

Metois and Wolf \cite{MeWo93} studied the structure of the Si(100)
surface with Reflection-Electron Microscopy (REM).  In the temperature
interval 1400-1455K the experiment shows a sudden increase in the
number of holes on the surface, and at 1455K the surface structure
cannot be resolved any longer and it is suggested that the surface at
this temperature becomes rough at the atomic scale.  In the
experimental study by Fraxedas {\it et.~al} \cite{FrFeCo94a} the
surface order of Si(100) was studied as a function of the surface
temperature by means of X-ray Photo-electron Diffraction (XPD).  At
1400 K the order suddenly decreases, and this is attributed to a phase
transition where the two first surface layers melt and form a liquid
layer. For the Si(100) surface molecular-dynamics simulations based on
the Stillinger-Weber potential have been used to study the solid-liquid
interface \cite{uzi}, but there are no studies of the surface melting.
To our knowledge, the only first principles molecular-dynamics study of the melting of a
semiconductor surface is by Takeuchi {\it et.\ al.} \cite{TaSeTo94}.
They study a Ge(111) surface at a temperature close to the melting
point and find evidence for the existence of an incomplete melted
phase.

 The experiment by Metois and Wolf shows
that  the Si(100) surface  has an increasing
concentration of vacancies as a function of the temperature. 
Since the vacancy formation takes place on a time scale of seconds, it
is not possible to study the vacancy formation in a molecular dynamics
simulations,  however, given a certain vacancy concentration we can
study what is the  surface structure attained within a
time scale of   $\approx 10-100$ ps. 

 In this report we present a molecular-dynamics study
of the Si(100) surface with and without vacancies
 for a number of temperatures
in the range 1100-1750K.   To describe the bonding of Si we have used
the Effective-Medium Tight-Binding (EMTB) model \cite{StChJaNo94b}. We
show that this approximate total energy method describes the
melting point and the properties of liquid Si very accurately. Using
the computational efficiency of the method to simulate a slab with a
unit cell containing 188 atoms for up to 35 ns at a number of
different temperatures and surfaces with and without defects, we show
that while the defect-free surface shows no sign of premelting, a
surface with 25\% of dimer vacancies shows melting of the first layer
100 K below the melting point. We show that the results of the
simulation can be used to understand the experimental observations for
this system.

\section{Technical Details}

The Effective-Medium Tight-Binding (EMTB) model \cite{StChJaNo94b} is based on
Effective-Medium Theory \cite{NoLa80,JaNoPu87} in which the
total energy calculations are simplified by comparisons to a reference
system(the effective medium). The so-called one-electron energy sum is
calculated using an 
LMTO tight-binding model \cite{AnJe84}.    The main approximations in
the  model is the  use of a first-order LMTO tight-binding model and a
non-self-consistent electron density \cite{ChJaNo91} and potential
\cite{StChJaNo94a}.  In previous studies we have demonstrated the
ability of the model to describe the low temperature structure of the
Si(100) surface,\cite{StChJaNo94b} and generally we have found the
accuracy for the total energy and atomic structure of solid Si systems
to be correct within 10-20 percent, which is comparable to plane-wave
calculations with a cutoff of 8 Ry.  However, the EMTB method is in
computational efficiency comparable to empirical tight-binding methods
and for the system sizes treated here something like three orders of
magnitude faster than 
plane-wave calculations. It is therefore very well suited for large
scale molecular-dynamics simulations.

The super-cell for the calculation is a 12 layer thick slab with 16
atoms in each layer except for the upper layer in which we have
introduced a vacancy concentration of 25 percent by removing four
atoms; thus a total of 188 atoms.  Since all atoms are free to move we
obtain information of a clean surface and a surface with vacancies in
the same simulation run.  The classical equations of motion for the
nuclei are integrated using
the Verlet algorithm with a time step of 1.08 fs and we control the
 temperature by using
 Langevin dynamics \cite{He86} with a friction
coefficient of 2 ps$^{-1}$. The lattice constant is determined by
scaling the EMTB (T=0K) value (10.167$a_0$) using the experimental
thermal-expansion coefficient.

\section{Results}
Initially, the atomic structure of the bottom surface consists of two
dimer rows with each 4 dimers, while the vacancies are introduced on
the upper surface by removing a dimer from each row. Starting from the
c(2x4) reconstruction we first increase the temperature of the
thermostat to 1100K. After thermalization this system is followed over
3ps. The final configuration is used as input for four other
simulation temperatures: 1450K, 1550K, 1650K, and 1750K, which are
followed over 29ps, 35ps, 33ps and 18ps, respectively.

We first summarize the calculations by showing in Fig.~1 the average
structure factor parallel to the dimer rows, $|S_{-11}|^2$, as a
function of the simulation temperature. The structure factor is
projected onto 3 intervals of the z axis corresponding to the position
of layers 1-2, 3-4, and 5-6 on both the vacancy covered and the clean surface.
The structure factor is averaged over the atomic configurations of the
last 2ps, 15ps, 15ps, 15ps and 5 ps for each of the five simulations
temperatures, respectively. The calculated variances in the structure
factors show that except for the 1650K calculation, the simulations
seem to have obtained an equilibrium state.  In Fig.~2 we show the
corresponding average atomic density perpendicular to the surfaces,
and for the 1550K calculation a snapshot of
 the atomic-structure is shown in Fig.~3.

The observed atomic structures obtained in the simulations
are the following: At 1100K the two surfaces show no tendency to break
up the dimer rows; each dimer stays in a buckling mode moving up and
down with a frequency of $\approx 3$ THz, and the buckling angle
reverses independently of the configurations of the other dimers. For
the clean surface we find this to be the picture up to 1750K where the
entire slab melts, and Fig.~1b shows that the structure factor for
this surface is constant up to the melting transition.

The vacancy surface shows a much more complex behaviour. At 1450K we
observe diffusion of one of the vacancies within the simulation time.
After the diffusion the dimers remain intact and the structure factor
shown in Fig.~1a is only slightly affected. At 1550K the two first
layers of the vacancy surface lose all their order and form a
disordered state, see Fig.~3, and  there is a strong interaction with
the third layer
which forms an interface between the bulk structure and  the
disordered surface. From Fig.~1a we see that now the
 structure factor has vanished for the two outermost layers, while the
structure factor of the third layer is slightly decreased due to the
interaction with the disordered state.  At 1650K the first four layers
disorders (cf. Fig.~1a and Fig.~2). At 1750 K all the layers melt
after a short transient period, and Fig.~2 shows a constant atomic
density $n=2.54g/cm^3$ in the range $-10a_0 < z < 10a_0$. This is an
increase of 6 percent compared to the crystalline density, and
 in good
agreement with the experimental atomic density $n=2.51g/cm^3$ for
liquid silicon at 1750K.

We first extract from the simulations the Si
bulk melting point, $T_m$, predicted by the EMTB model. At 1750K the
whole slab melts, without a partially-melted transient period, and we
therefore have $T_m < 1750K$. At 1550K there are two phases present in
the calculation with a large interaction between the two.  Since
Fig.~1a shows that there are only small fluctuations in the structure
factor of the layers, this strongly suggests that the simulation has
obtained an equilibrium state. Furthermore, if the melting temperature
was close to 1550K, the temperature in the 1650K simulation would be
well above the melting point and the slab should readily melt since a
melted phase is already present in the calculation. From these
considerations we estimate a melting temperature of $T_m \approx$
1650$\pm 100$K, in good agreement with the experimental melting point
of 1685K.

The rather large error bar in the melting temperature comes from
finite size effects leading to large fluctuations in the temperature.
On top of this there may be a systematic overestimate of the melting
temperature in the simulation because of the fixed super-cell size
parallel to the slab. However, the possibility of the system to
respond with volume changes perpendicular to the slab reduces this
problem considerably. The presence of the surface thus allows the
liquid to acquire the correct density as noted above.

In the simulation with the thermostat set at 1650K the temperature
fluctuates around the melting point, and we may have a coexistence of
solid and liquid phases at this temperature. This suggests that the
melting of the two additional layers in the 1650K simulation is rather
due to large fluctuations of the temperature around the melting point
than the existence of a surface phase transition where four layers
disorders. Note also the large fluctuations observed in the structure
factor of layer 5-6 at this temperature, shown in Fig.~1a, which makes
it questionable whether an equilibrium state has been obtained within
the simulation time.

The temperature in the 1550K simulation, on the other hand, is well
below the bulk melting point and the atomic structure in this case
must be due to a surface phase transition taking place at a
temperature $T_F < 1550$. From simulations of metal surfaces it is
known \cite{StNo} that just at the surface melting temperature the
defects become mobile, and we therefore estimate $T_F = 1450\pm 100K$.
Since we have $T_F < 1550 < T_M$ we can expect to have obtained
equilibrium within our simulation time for this temperature. We now
consider the 1550K simulation in more detail.

In Fig.~4 we show the radial distribution function averaged over the
atoms in layer 2 and layer 6-7. For comparison we also show similar
averages obtained for the 1750K simulation in which case the whole
slab has become liquid. The figure clearly shows that the two
outermost layers in the 1550K simulation have formed a liquid state,
with an average coordination very different from the bulk state, but
almost identical to that of liquid Si. It is worth noting that the
radial distribution function obtained for the liquid Si is in good
agreement with experiment \cite{experiment} and that obtained by Stich
{\it et. al.} \cite{StCaPa88} in a Car-Parrinello calculation (the
position of the first maximum is $4.67a_0 (4.65a_0)$ and integration
up to $5.90 a_0$ gives an average coordination of 6.55(6.5), where the
values obtained in Ref.~\cite{StCaPa88} are given in parenthesis).
The formation of a liquid surface bilayer
 in the 1550K simulation is 
further illustrated by
Fig.~5, where we from the lateral mean-square displacement  of the
surface atoms deduce a two
dimensional diffusion constant of $D=2.1
\times 10^{-5}$ cm$^2$/s. 
We also observe a jump in the average potential energy during melting at
1550K from which we deduce a latent heat of melting the surface atoms
of $\Delta E \approx 3 eV$. This must be compensated by an increase
in the entropy $\Delta E = N k_B T\Delta S$ (where $N$ is the number
of atoms taking part in the phase transition), which when attributed
to the atoms in the two first layers ($N=28$), gives an increase in
the entropy of $\Delta S \approx 0.8$.

We now relate our simulations for the Si(100)  to those of
Ref.~\cite{TaSeTo94} for the Ge(111) surface. The outermost layer of
the Ge(111) surface consists of fourth a monolayer Ge adatoms, which
gives this surface  a more open structure than the Si(100)
surface. The Ge(111) surface has more resemblance with the  vacancy
covered Si(100) surface, and there are indications that the adatoms on the
Ge(111) surface and the vacancies on the Si(100) surface play similar
roles in the surface melting process. Similar to the vacancy diffusion
we found in the 1450 K simulation, Takeuchi {\it et. al.} found a
metastable state in the first part of their simulation  where only the
adatoms diffuse. Afterwards the Ge(111)
surface undergoes a phase transition where the two first layers become
liquid, and the increase in entropy and the
diffusion constant are very similar to the values we have found for
the vacancy covered Si(100) surface. (For Ge(111) the
values are $D_{Ge}=3.5 \times 10^{-5}$ cm$^2/s$, $ \Delta S_{Ge}
\approx 1.0$). 

For the Ge(111) surface
 it was also found that the electronic structure of the melted
layers was metallic, and lately this has been supported by an
 electron energy loss spectroscopy (EELS) study of the surface
conductivity of Ge(111)\cite{MoDhSaSaPeTo94}. However, we only find a
weak metallic  behaviour for the incomplete 
melted Si(100) surface. This is illustrated in Fig.~6, which shows the
Projected Density of States(PDOS) of the atoms in the two first 
layers of the vacancy surface(solid line) and the clean surface(dotted
line), and  we see that the number of states in the gap is only slightly
increased for the incomplete melted vacancy surface compared to the non-melted
clean surface. We therefore expect  that  the surface melting of
Si(100) will  be difficult to
detect with EELS.

\section{Comparison with experiments}
We now compare our findings to the experimental results of
Ref.~\cite{MeWo93,FrFeCo94a}. One of our  results is
that the phase transition only takes place when a high concentration
of vacancies is present on the surface, and the question arises what
are the vacancy concentrations in the experiments. STM studies yield
estimates that the surface at room temperature has a vacancy
concentration of 5-10 percent.\cite{HaTrDe86} When the surface is
heated up one may observe atoms evaporating, and since they leave a
vacancy behind this process may increase the vacancy concentration.
Metois and Wolf found that at temperatures below 1400K the sublimation
process proceeds slowly enough, that the sublimation holes are filled
with atoms diffusing from step edges and the vacancy concentration is
relatively stable up to this temperature.
In the temperature interval 1400-1455K the number of holes increases
drastically and  at 1455K the surface structure cannot be resolved any
longer.

The experiment by Metois and Wolf 
is in a non-equilibrium state and  they relate the observed behaviour to
kinetic surface roughening.  The kinetic roughening is caused by the larger
activation energy for sublimation compared to the activation energy for
diffusion, which means that  the sublimation rate will increase faster
with temperature
than the diffusion rate, and at some temperature 
the sublimation rate will become too fast for the vacancies to be filled with
atoms diffusing from step edges. However, we note that the observed
behaviour is not necessarily due to kinetic roughening but can alternatively
 be due to the existence of a roughening phase transition. Both
kinetic roughening and a roughening phase transition will give rise to an increased
vacancy concentration as a function of the temperature,
and the vacancy concentration in the experiments of
Ref.~\cite{MeWo93,FrFeCo94a} at 1450K  will therefore be well above the 5-10
percent concentration at room temperature. From the available data we cannot determine the exact vacancy
concentration in the two experiments, and  probably they will not have
the same 
concentrations, since it will depend on experimental
details, such as base pressure and the crystal miscut angle.

It is not possible to use molecular dynamics simulations to determine
the (equilibrium or non-equilibrium) concentrations of vacancies since
the simulation time scales are too short. However, for a given vacancy
concentration molecular dynamics can be used to study the attained
surface structure within a time scale of   $\approx 10-100$ ps, which
is long compared to fundamental surface vibration frequencies. In the
simulation we have studied a vacancy concentration of 25 percent,
where we find a phase transition at 1550 K to an incomplete melted
state with two melted layers. We expect the same phenomena to take
place for higher or slightly lower vacancy concentrations. 
 However,  the results cannot be generalized to
low vacancy concentrations, since in this case 
 the time
scale of the formation of an incomplete melted phase may
 become comparable to the time scale of the processes that 
 alter the vacancy concentration.
For instance, on the time scale of desorption of Si dimers (seconds)
the simulation for the clean surface will  show incomplete
melting, since the desorption of Si dimers 
will give rise to 
 an increased vacancy concentration  on the surface. In a real
experiment there will also be adatoms diffusing of step edges and
 adsorption of atoms from the gas phase, and these processes 
 will lower the  vacancy concentration.

We now make quantitative comparisons with the experimental study of
Ref.~\cite{FrFeCo94a}, where the XPD anisotropy ($A_{(hkl)}(T)$) is
measured along the (100) and (111) directions as a function of
temperature. The anisotropy in a given direction is defined as
$A_{(hkl)}(T) =
(I_{max}(T)-I_{min}(T))/(I_{max}(300K)-I_{min}(300K))$, where
$I_{max}$ ($I_{min}$) stands for the maximum (minimum) intensity for a
given peak. We have simulated the XPD intensity peaks using
Single-Scattering Cluster (SSC) calculations \cite{Fa84}, with the
same SSC parameters as quoted in Ref.~\cite{FrFeCo94b}.  For the
electron mean-free path the value is 25\AA \ (corresponding to the
energy of the 2s photo electrons), in spite of the generally accepted
rule of using half of its nominal value. \cite{NaStGrOsSc93} The
anisotropy along the (111) direction is extremely sensitive to the
actual choice of mean-free path, and we have therefore chosen only to
make comparison with the anisotropy along the (100) direction, which
we have found to be rather insensitive to the value of the mean-free
path.

For each simulation temperature we have calculated the anisotropy
along the (100) direction for 20 different configurations and Fig.~7
shows the corresponding averages.  We have performed calculations
using both a mean-free path of 12.5\AA \ and 25\AA . In each case the
anisotropies have been scaled such that at 1100K the anisotropy of the
clean surface fits the experimental measurements, and Fig.~7 shows
that the relative anisotropies are quite insensitive to the choice of
mean-free path. We may therefore relate our calculations directly to
the experimental measurements.  First of all we see that vacancies
lead to a drop in the anisotropy, and the observed increase in the
vacancy concentration of Ref.~\cite{MeWo93} should therefore be
measurerable by XPD. We find that an increase in the vacancy
concentration by 10 percent would be consistent with the small
decrease in the experimental anisotropy observed in the temperature
range $\approx 1390-1425K$. Since we must expect \cite{HaTrDe86} that
the surface already has a vacancy concentration around 10 percent at
1390K, the vacancy concentration will be 20 percent at 1425K, very
similar to the concentration in our simulation.

An increase in the vacancy concentration, however, can hardly describe the
overall drop in the anisotropy observed in the range 1390-1500K.
Fig.~7 indicates strongly that the decrease in the anisotropy after
1425K is related to the defect-induced disordering we have described
above.  This picture is in accordance with Ref.~\cite{FrFeCo94b},
where they found that the experimentally observed drop in anisotropy
could be explained by assuming a disordered bilayer at the surface.
The experiment also shows that the thickness of the melted layer is
constant upto the bulk melting point, again indicating that the
melting of the two additional layers in the 1650K simulation is
related to bulk melting rather than surface melting.

\section{summary}
In summary, we have performed molecular-dynamics simulations for the
Si(100) surface using the EMTB model.  We find that in the presence of
a vacancy concentration of 25 percent the surface undergoes a phase
transition at 1550K, where the two first layers melt and form a liquid
state. The incompletely melted state shows several similarities to the
incomplete melting of Ge(111)\cite{TaSeTo94}, except that the
electronic surface 
structure shows a much weaker metallic behaviour.
 We have also shown that the simulation results are in both
quantitative and qualitative agreement with experimental studies of
the surface structure.

\acknowledgements
We thank C. Fadley for providing us the program for the SSC
calculations and J. Fraxedas for instructions on how to use the
program. Many discussions with P. Stoltze and E. Tosatti are
gratefully acknowledged.  The Center for Atomic-scale Materials
Physics is sponsored by the Danish National Research Foundation.
 Kurt Stokbro acknowledges eec contract ERBCHBGCT
920180 and contract ERBCHRXCT 930342 and  CNR project   Supaltemp.

\begin{figure}
\caption{The  equilibrium average structure factor parallel to the dimer rows,
  $|S_{-11}|^2$, as a function of the simulation temperature.  The
  structure factor is projected onto 3 intervals of the z axis
  corresponding to the position of layer 1-2, 3-4, and 5-6, for a) The
  vacancy covered surface, and  b) the clean surface.}
\end{figure}

\begin{figure}
\caption{The  equilibrium average atomic density
  perpendicular to the surfaces for each simulation temperature.}
\end{figure}

\begin{figure}
\caption{A snapshot of the atomic structure in the 1550K simulation
at $t=30$ps. The supercell is repeated twice in the horizontal direction.}
\end{figure}

\begin{figure}
\caption{The radial distribution function projected onto layer 2 and
  layer 6-7 for the 1550K simulation(solid line) and the 1750K
  simulation (dashed line).}
\end{figure}

\begin{figure}
\caption{ The atomic displacement in the surface plane projected onto
layer 1-2, 3-4, and 5-6,  as a function of time for the 1550K simulation.  
The 2-dimensional
diffusion constant, $D$, for layer 1-2 is defined by $<r^2> = 4 D t$.}
\end{figure}

\begin{figure}
\caption{ The PDOS of layer 1-2 of the vacancy surface(solid line) and the clean
surface (dotted line) in the  1550K simulation. The average is taken
over the last 15 ps of the simulation.}
\end{figure}

\begin{figure}
\caption{ The XPD
  anisotropy as obtained from SSC calculations with an electron
  mean-free path of 12.5\AA \ (solid line) and 25 \AA \ (dashed line)
  as a function of the simulation temperature. The experimental
  measurements are from Ref.~\protect\cite{FrFeCo94a}.}
\end{figure}

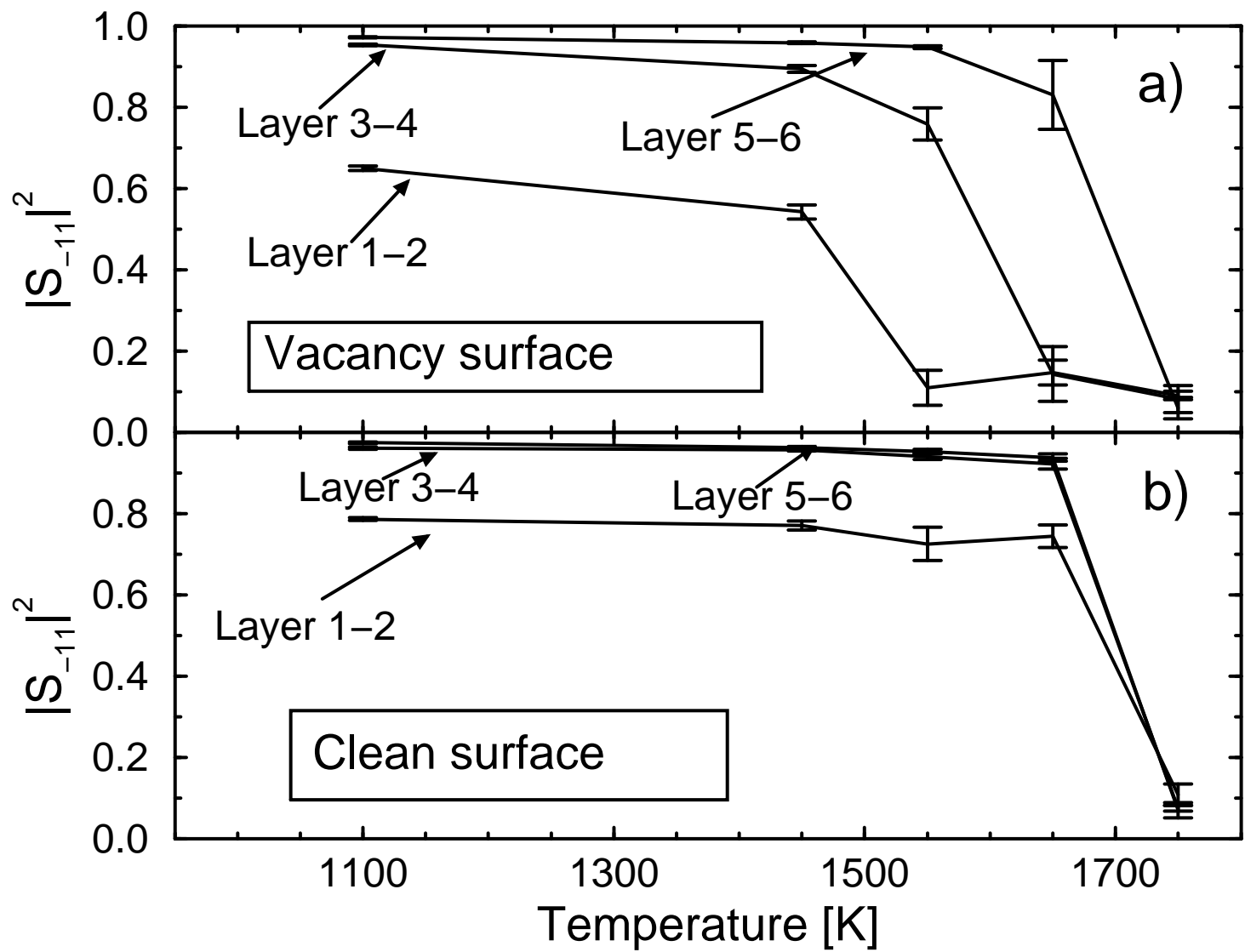
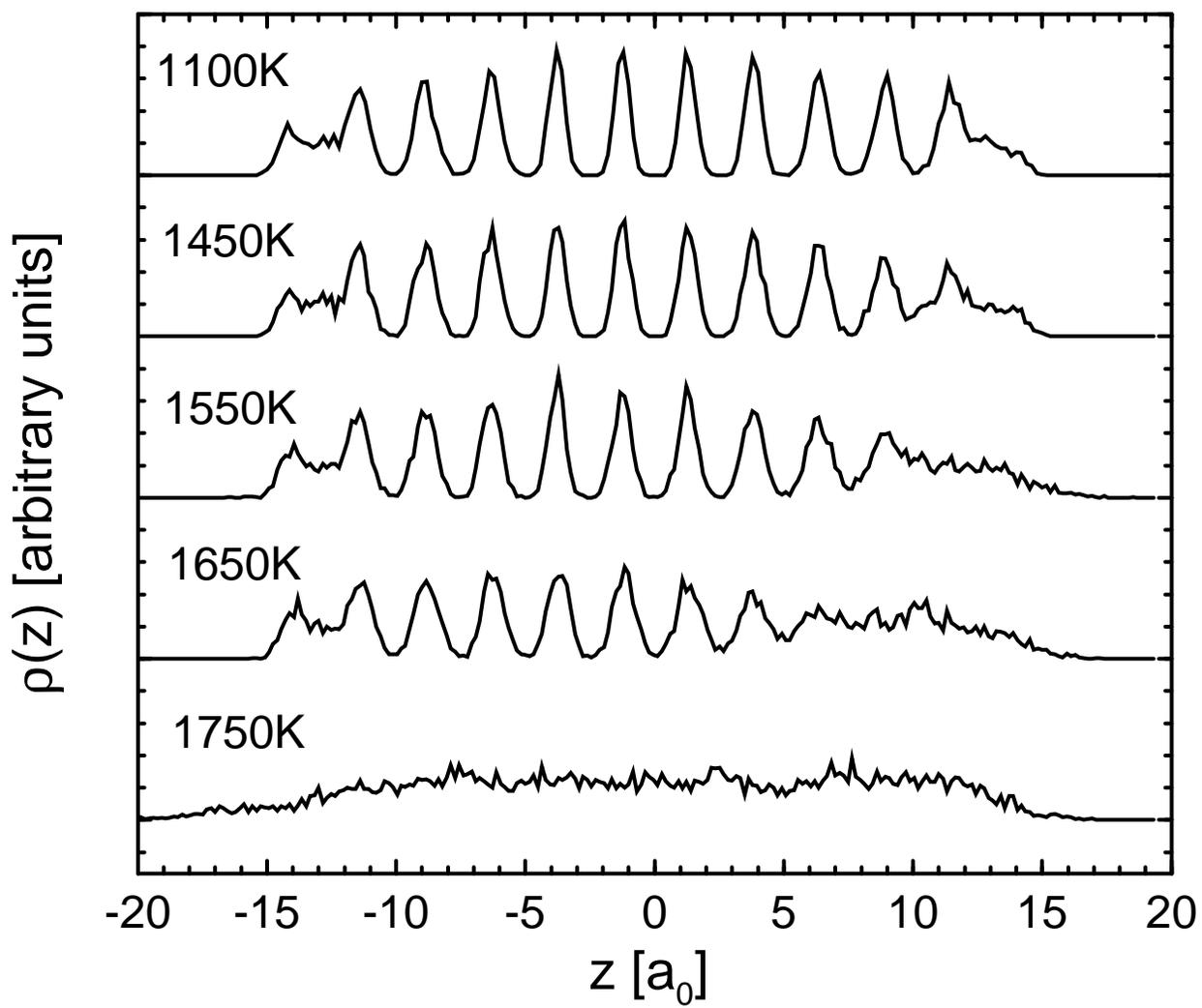
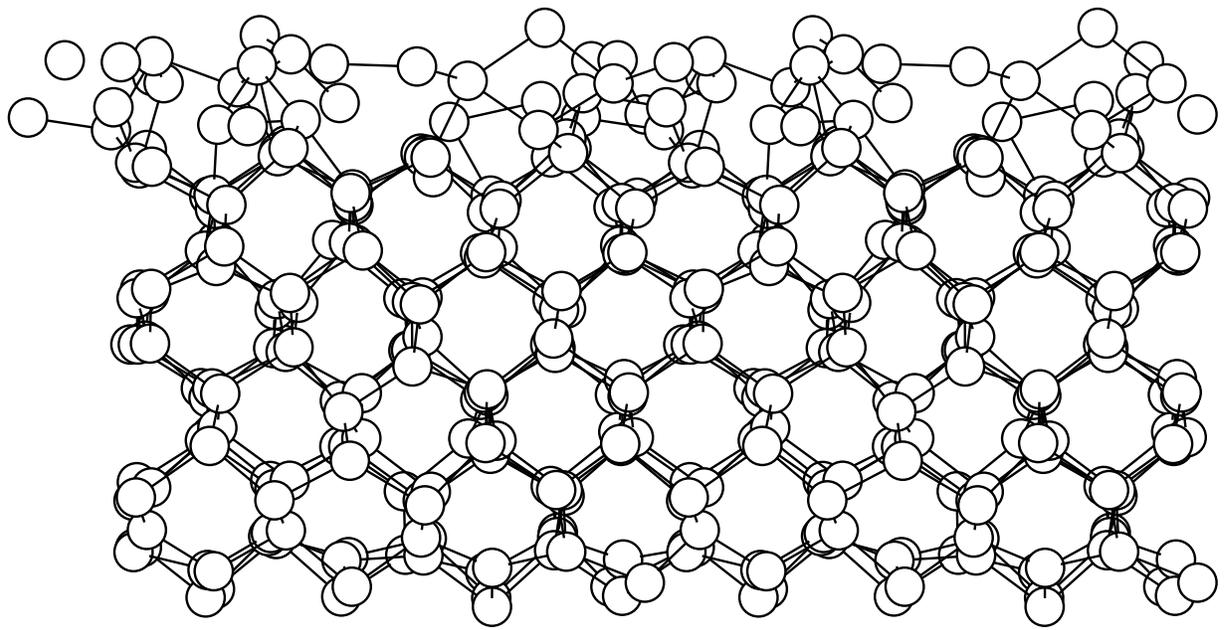
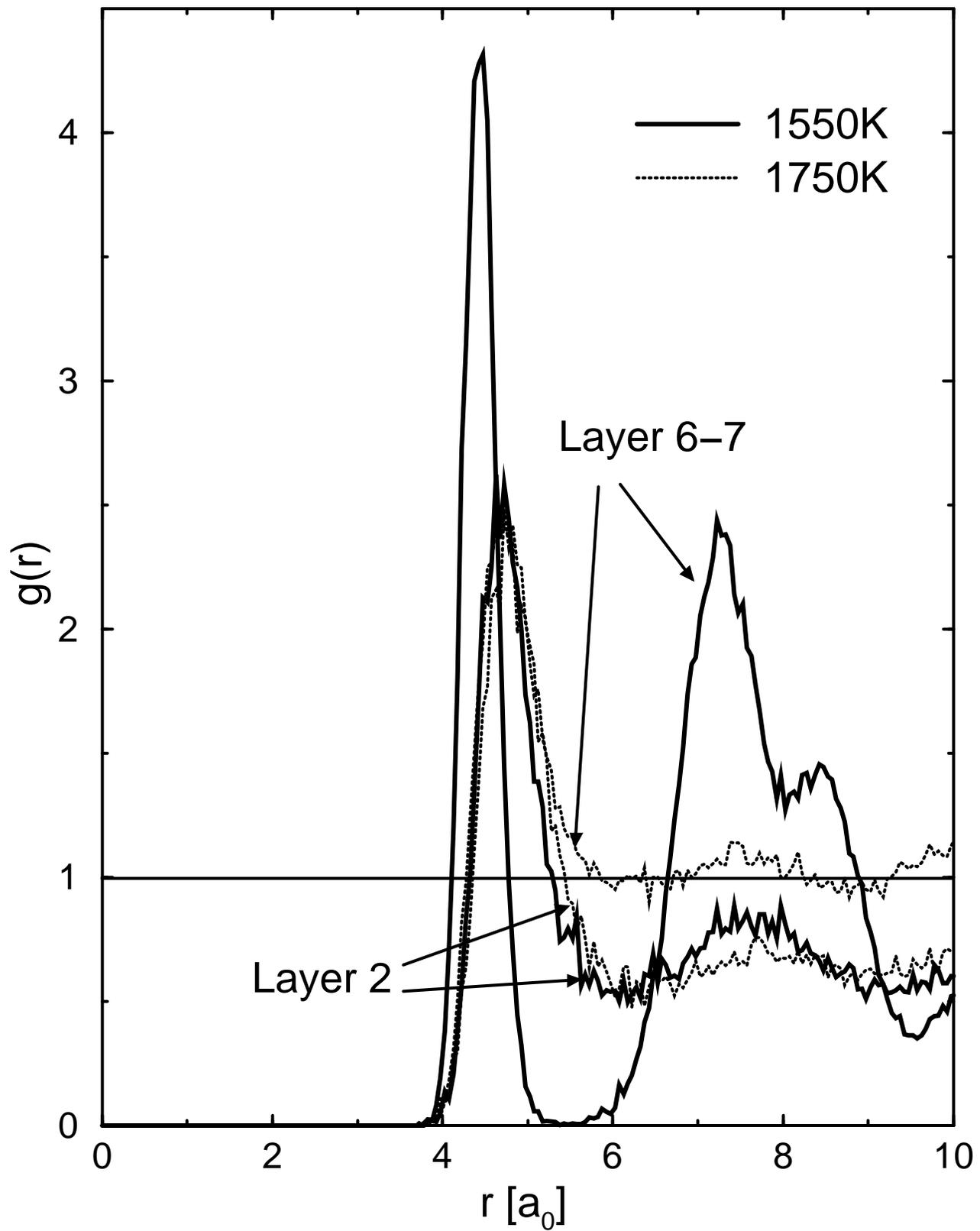
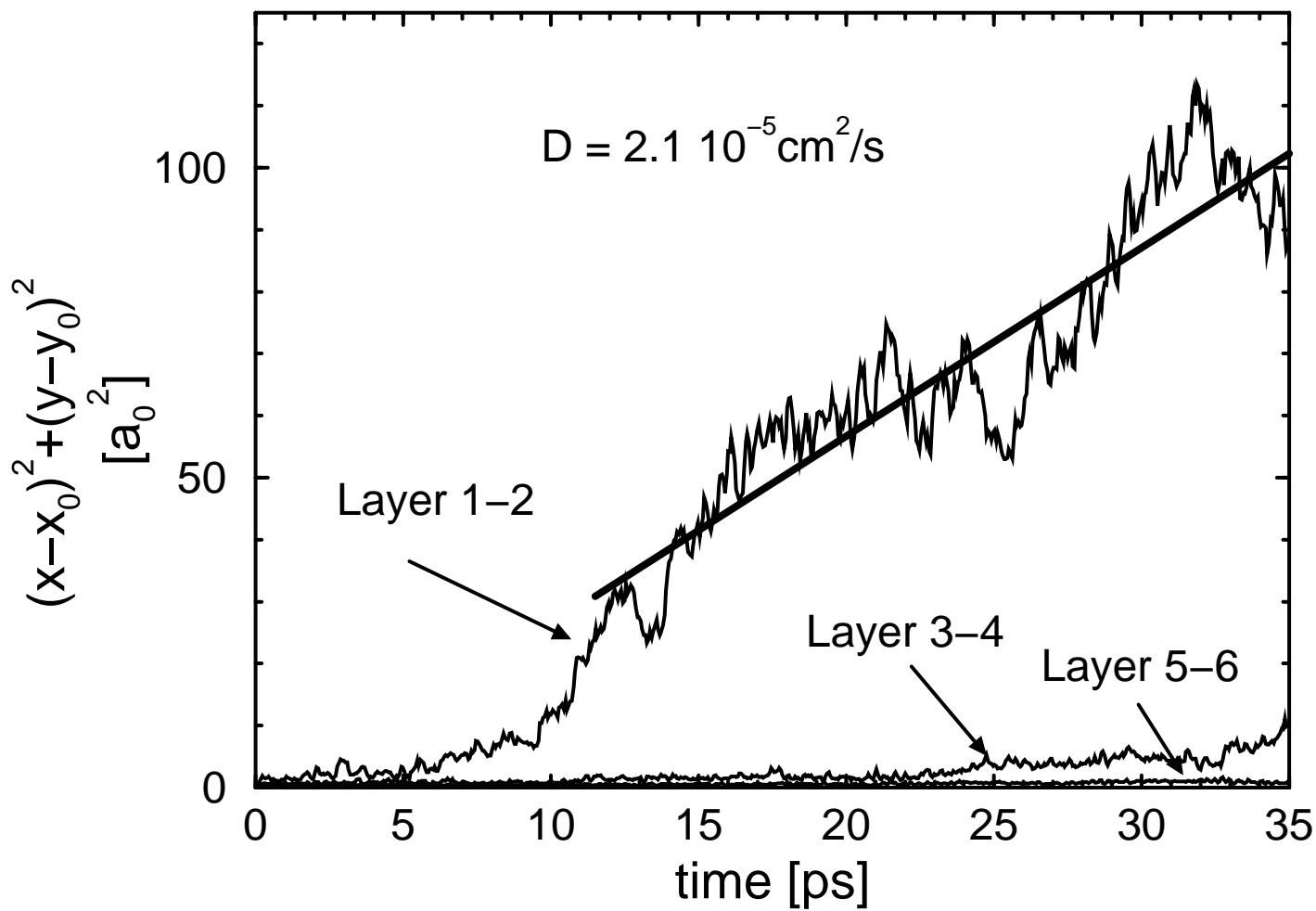
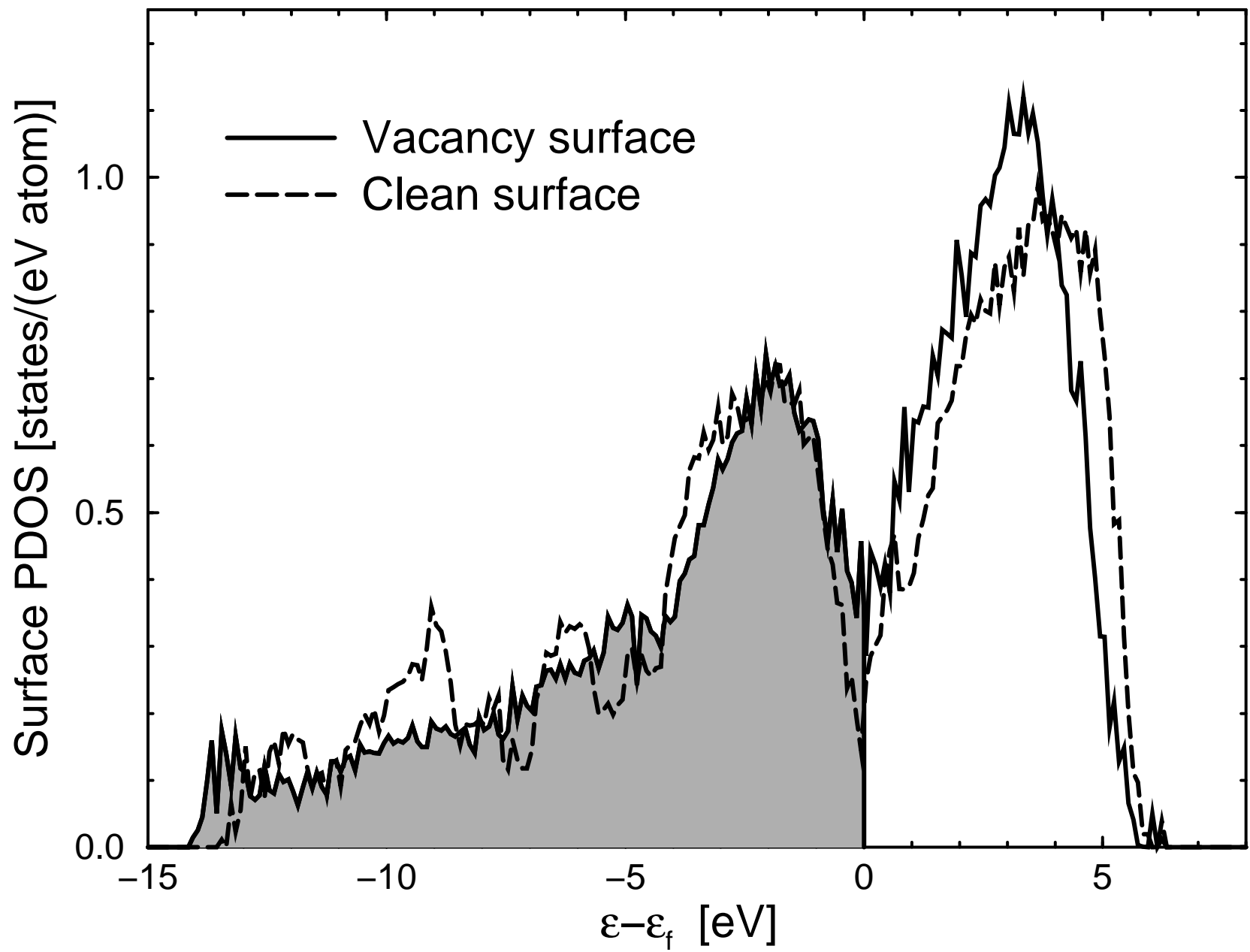
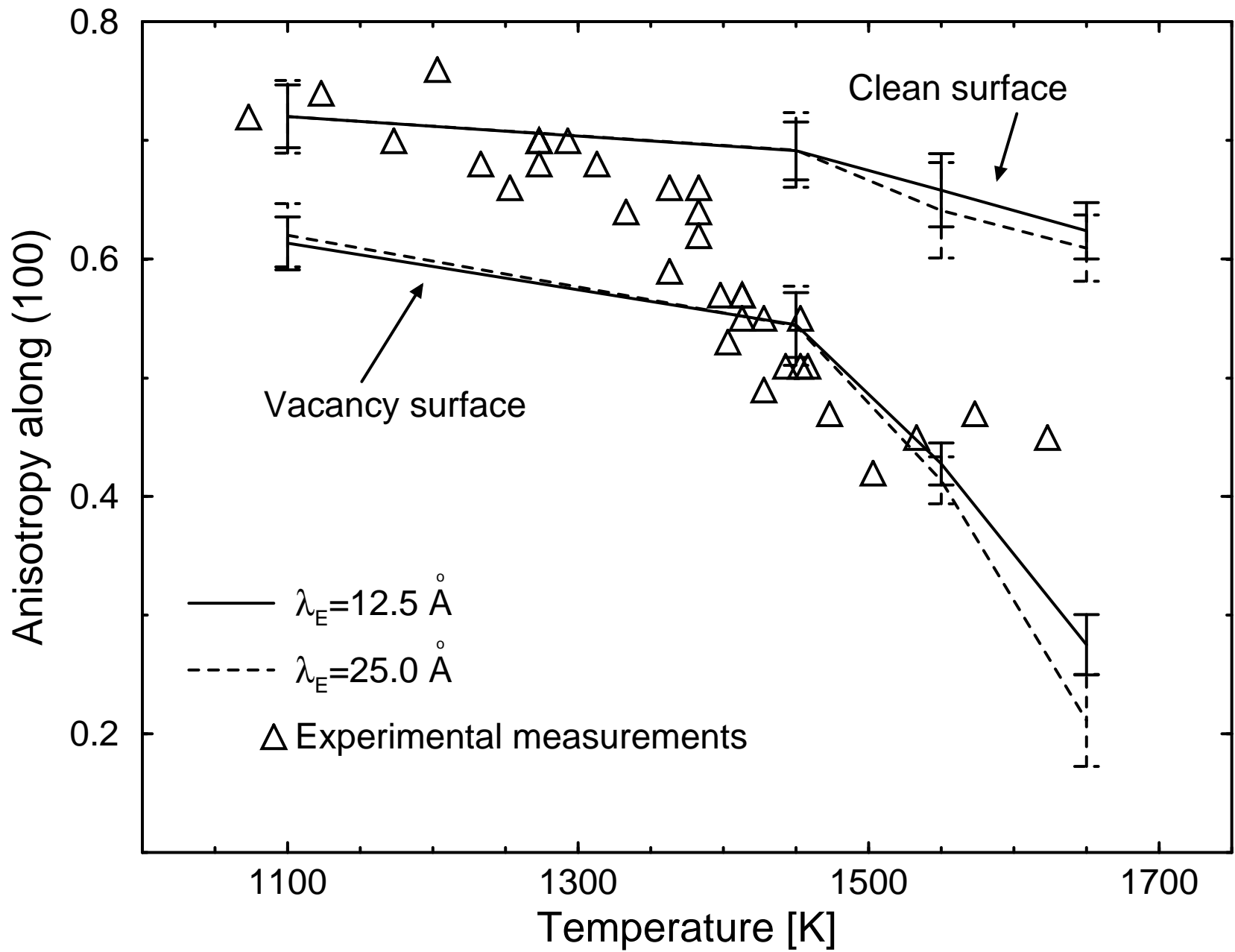
\end{document}